\documentclass[english,pra,twocolumn,superscriptaddress]{revtex4-2}
\usepackage[T1]{fontenc}
\usepackage[latin9]{inputenc}
\usepackage{color}
\usepackage{array}
\usepackage{mathrsfs}
\usepackage{amsmath}
\usepackage{amssymb}
\usepackage{graphicx}

\makeatletter

\providecommand{\tabularnewline}{\\}

\usepackage{times}
\usepackage{array}
\usepackage{color,soul}

\makeatother

\usepackage{babel}
\begin{document}
\title{Entanglement and classical nonseparability convertible from orthogonal
polarizations}
\author{Minghui Li}
\affiliation{Institute of Applied Physics and Materials Engineering, University
of Macau, Macau S.A.R., China}
\author{Wei Wang}
\affiliation{Institute of Applied Physics and Materials Engineering, University
of Macau, Macau S.A.R., China}
\author{Zikang Tang}
\email{zktang@um.edu.mo}

\affiliation{Institute of Applied Physics and Materials Engineering, University
of Macau, Macau S.A.R., China}
\author{Hou Ian}
\email{houian@um.edu.mo}

\affiliation{Institute of Applied Physics and Materials Engineering, University
of Macau, Macau S.A.R., China}
\begin{abstract}
The nonclassicality of a macroscopic single-mode optical superposition
state is potentially convertible into entanglement, when the state
is mixed with the vacuum on a beam splitter. Considering light beams
with polarization degree of freedom in Euclidean space as coherent
product states in a bipartite Hilbert space, we propose a method to
convert t\textcolor{black}{he two orthogonal polarizations i}nto simultaneous
entanglement and classical nonseparability through nonclassicality
in the superpositions of coherent and displaced Fock states. Equivalent
Bell state emerges from the resulted superpositions and the proportion
of mixed entanglement and nonseparablity is determined by the displacement
amplitudes along the polarization directions. We characterize the
state nonclassicality via features in Wigner distributions and propose
an experimental method for generating these states and measuring them
via homodyne tomography.
\end{abstract}
\maketitle

\section{Introduction}

At one end in the realm of optics, light of quantum nature is understood
through the corpuscular concept of photon written as a Fock state.
At the other end, its classical counterpart is embodied by a monochromatic
beam written as a coherent state. The linkage between the two ends
is the extension of finite-dimensional state space to infinite dimensions,
in which the coherent state is equivalent to an infinite superposition
of Fock states with Poisson distributed photon statistics~\citep{Loudon}.
Although the delineation between the classical regime and the quantum
regime remains blurred, it is quite clear that the progression from
the quantum state to the classical state of light is nuanced and a
class of states belonging to neither extreme, known as nonclassical
states of light~\citep{Vogel_00,Lee,DODONOV}, occupies the middle
ground.

The inclusive term comprises a variety of states. Besides the familiar
squeezed states and Schrödinger cat states, it also encompasses a
genre of states connected to the nominally classical end: derivative
states from coherent states. It includes the single-photon-added coherent
state (SPACS, essentially $a^{\dagger}\left|\alpha\right\rangle $)~\citep{Agarwal-1,Zavatta},
the displaced Fock state (DFS, $D(\alpha)\left|1\right\rangle $)~\citep{Lvovsky-1,Notation},
and superpositions of two coherent states ($\left|\alpha\right\rangle +\left|\beta\right\rangle $)~\citep{Schleich}.
The forementioned states have a quantifiable nonclassicality determined
by entanglement potential~\citep{asboth,Killoran}, which measures
the entanglement convertible from them when mixed with a vacuum state
using just linear optical components and photodetectors. These investigations
open up the possibility of quantum optical computation using macroscopic
nonclassical states derived from classical light beams~\citep{Bruno,Lvovsky,Sychev,Biagi}.
However, to what exact quantum state an arbitrarily polarized laser
beam is mappable remains an open question. Specifically, from a quantitative
perspective, it is still unclear how quantum and classical characteristics
vary when two macroscopic orthogonal polarizations are converted into
entangled states.

To these ends, we study entanglement as well as classical nonseparability~\citep{Spreeuw,Gabriel,Qian,Karimi,Rafsanjani}
obtainable from an plane-wave electric field with two orthogonal polarizations.
The latter has seen recent usages in coding quantum-like information~\citep{Milione,Pierangeli,Shen,Qing}.
To retain the full gradation over the progression from the classical
to the quantum regime, we consider the electric field that experiences
several stages of splitting and recombining in a light path, where
the only quantum component is a quadrature operation realizable with
parametric down conversion and projective detection. We show that
the full range of entanglement and classical nonseparability can be
obtained and co-exist with suitable polarizations in the two directions
when converted. The convertibility demonstrates not only the potential~\citep{asboth}
that entanglement would emerge from classical states, but also the
overlap of classical nonseparability originating from two macroscopic
field components. 

Here, entanglement and classical nonseparability are treated in a
unified Hilbert space for quantifying their convertibility from the
macroscopic polarizations. Specifically, the $x$ and $y$ polarizations
of an electric field vector in Euclidean space $\mathbb{R}^{3}$ are
expressed as coherent states in the infinite dimensional Hilbert spaces
$\mathcal{H}_{x}$ or $\mathcal{H}_{y}$ parametrized by continuous
displacements. The entanglement and nonseparability then both emerge
as measures on a superposition state $\left|\psi\right\rangle $ in
the product space $\mathcal{H}=\mathcal{H}_{x}\otimes\mathcal{H}_{y}$.
As quantifying metrics, their distinction is only mathematical: entanglement
appears as a functional (we measure it in negativity~\citep{Vidal})
directly on the Hilbert space vector $\left|\psi\right\rangle $ while
nonseparability appears as a functional (we measure it in Schmidt
number~\citep{Gabriel,Qian}) on the Euclidean space vector $\mathbf{E}=\langle\psi|\hat{E}|\psi\rangle$
derived from the field operator $\hat{E}$.

This unified approach assists in distinguishing the quantum entanglement
and classical nonseparability within a single nonclassical state of
light, according to the intuition to distinguish packetized photons
expressed as Fock states from classical single-mode beams expressed
as coherent states. For examples, as our discussions below will show,
the apparent product state $\left|\psi\right\rangle =\left|\alpha\right\rangle _{x}\otimes\left|i\alpha\right\rangle _{y}$
has a maximal Schmidt number for classical nonseparability but zero
negativity for entanglement. At the opposite extreme, the superposition
$|1^{(\gamma)}\rangle_{x}|\delta\rangle_{y}+|\gamma\rangle_{x}|1^{(\delta)}\rangle_{y}$
of the orthogonal coherent states and DFSs, becoming the analogue
of Bell states in the continuous space, obtains maximal negativity
with vanishing Schmidt number. In these nonclassical states, the complex
displacements $\alpha$, $\gamma$, and $\delta$ derived from the
polarization amplitudes serve as important indicators for the eventually
convertible entanglement and classical nonseparability.

We give a detailed analysis of this convertibility when only one quantum
operation is inserted along a linear optical path below. We remark
that the choice of entanglement and classical nonseparability measures
are independent of the results obtained. For example, if we use the
Vogel-Sperling version of Schmidt number~\citep{Vogel} that measures
entanglement as a functional on $\left|\psi\right\rangle $, the obtained
variation against the displacements would coincide with those obtained
from the negativity. Wigner distributions representing the density
matrix of the quantum states are used throughout to visualize the
appearance of nonclassical states against the classical ones.

\section{Generating entangled state}

\begin{figure}
\begin{centering}
\includegraphics[width=8.5cm]{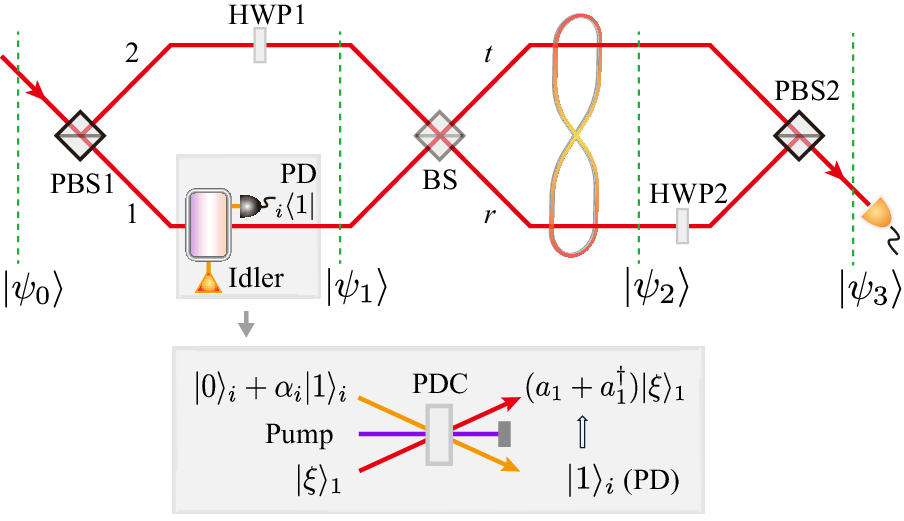}
\par\end{centering}
\caption{Gedanken scheme for constructing a macroscopic entangled state: an
initial product state $|\psi_{0}\rangle$ of a polarized beam  goes
through multiple stages to become entangled in the final state $\left|\psi_{3}\right\rangle $
before it is detected. First, $\left|\psi_{0}\right\rangle $ is split
into two paths by a polarizing beam splitter (PBS1), after which one
polarization branch (path 1) undergoes a conditional quadrature operation
$q=a+a^{\dagger}$. Shown in the inset, this operation is implemented
by a parametric down-conversion (PDC) followed by a photon detection
(PD), forming a superposition of a coherent state and a SPACS. The
other path 2 is rotated by a half-wave plate (HWP1) to interfere with
path 1 through a balanced beam splitter (BS). The branched beams described
by the product state $\left|\psi_{1}\right\rangle $ is thus mixed
to generate the entangled state $\left|\psi_{2}\right\rangle $ that
propagates along the transmission $t$ and the reflection $r$ directions
after splitting. The $r$ mode is rotated by HWP2 to align its polarization
orthogonal to the $t$ mode before they are recombined by PBS2 to
form the final macroscopic entangled state $|\psi_{3}\rangle$.~\protect\label{fig:scheme}}
\end{figure}

To substantiate the brief description above, we consider a gedanken
experiment that produces any entangled state of varying degrees of
negativity and Schmidt number, as shown in Fig~\ref{fig:scheme}.
A single-mode laser operating well above the threshold is considered
the input source. It shows a coherent state excitation and thus exhibits
an almost classical behavior~\citep{Loudon}. Regarded as a plane
wave carrying two independent polarizations, it has the joint state
$|\psi_{0}\rangle=|\xi\rangle_{x}|\eta\rangle_{y}\in\mathcal{H}_{x}\otimes\mathcal{H}_{y}$
where $x$ and $y$ indicate a pair of orthogonal polarization directions.
In other words, photons in $\mathcal{H}_{x}$ possess polarization
opposite to those in $\mathcal{H}_{y}$~\citep{Jauch}.

The laser beam is first split into two according to the polarization,
after which the $x$-polarized part along path 1 undergoes a quadrature
operation $q=a+a^{\dagger}$ while the $y$-polarization along path
2 is rotated by the HWP1 into $x$-polarization. The mathematical
$q$-operation is optically implemented by parametric down conversion
(PDC) with parametric gain $g$: the $\left|\xi\right\rangle _{1}$
beam serves as the signal while a beam with state $\left|0\right\rangle _{i}+g\xi|1\rangle_{i}$
serves as the idler input. At the output end, when the weak idler
is measured by a photon detector and only single-photon events are
selected, the signal output becomes conditioned and post-selected
to the pure state $(a_{1}+a_{1}^{\dagger})\left|\xi\right\rangle _{1}$,
a superposition of coherent and SPAC states~\citep{Zavatta,Resch}.
The nonclassicality of this superposition state is crucial for generating
quantum entanglement with a beam splitter~\citep{Kim02}, as we will
see later. 

At this stage, the system state is $\left|\psi_{1}\right\rangle =(a_{1}+a_{1}^{\dagger})\left|\xi\right\rangle _{1}\left|\eta\right\rangle _{2}/\sqrt{N}$
with $N$ being the normalization constant, where we have used the
path subscripts to differentiate the branched beams since both paths
are now $x$-polarized. The two beams are combined by a 50-50 beam
splitter (BS), which effectively perform the transformation $a_{1,2}\to(a_{t}\mp a_{r})/\sqrt{2}$
on the annihilation operators as well as their Hermitian conjugates
before and after the beam splitting~\citep{Prasad}. That means,
for instance, removing one photon in path 1 is equivalent to removing
one photon either at the transmission beam or the reflection beam.
Carrying out the algebraic operation on $\left|\psi_{1}\right\rangle $,
the BS output is the state
\begin{equation}
\left|\psi_{2}\right\rangle =\frac{(q_{t}-q_{r})}{\sqrt{2N}}\left|\frac{\eta+\xi}{\sqrt{2}}\right\rangle _{t}\left|\frac{\eta-\xi}{\sqrt{2}}\right\rangle _{r}\label{eq:psi2}
\end{equation}
in the product space $\mathcal{H}=\mathcal{H}_{t}\otimes\mathcal{H}_{r}$
of the two branches. The state $\left|\psi_{2}\right\rangle $ is
already an macroscopic entangled state with respect to the $t$ and
$r$ beams, converted from the nonclassical state $|\psi_{1}\rangle$.
The beam splitting effects the transformation $q_{1}\to(q_{t}-q_{r})/\sqrt{2}$
on the quadrature operator and the transformation $D_{1}(\xi)D_{2}(\eta)\to D_{t}((\eta+\xi)/\sqrt{2})D_{r}((\eta-\xi)/\sqrt{2})$
on the displacement operators (Cf. Appendix A). In the latter, the
displacement amounts from the vacuum are essentially the polarizations
in the macroscopic beam, i.e. the transmission and the reflection
contains polarization originated from both path 1 and path 2. Therefore,
measuring either the $t$- or the $r$-subspace would already yield
information from both the original $\mathcal{H}_{x}$ and $\mathcal{H}_{y}$
spaces.

\begin{figure}
\begin{centering}
\includegraphics[bb=0bp 0bp 474bp 500bp,clip,width=8.5cm]{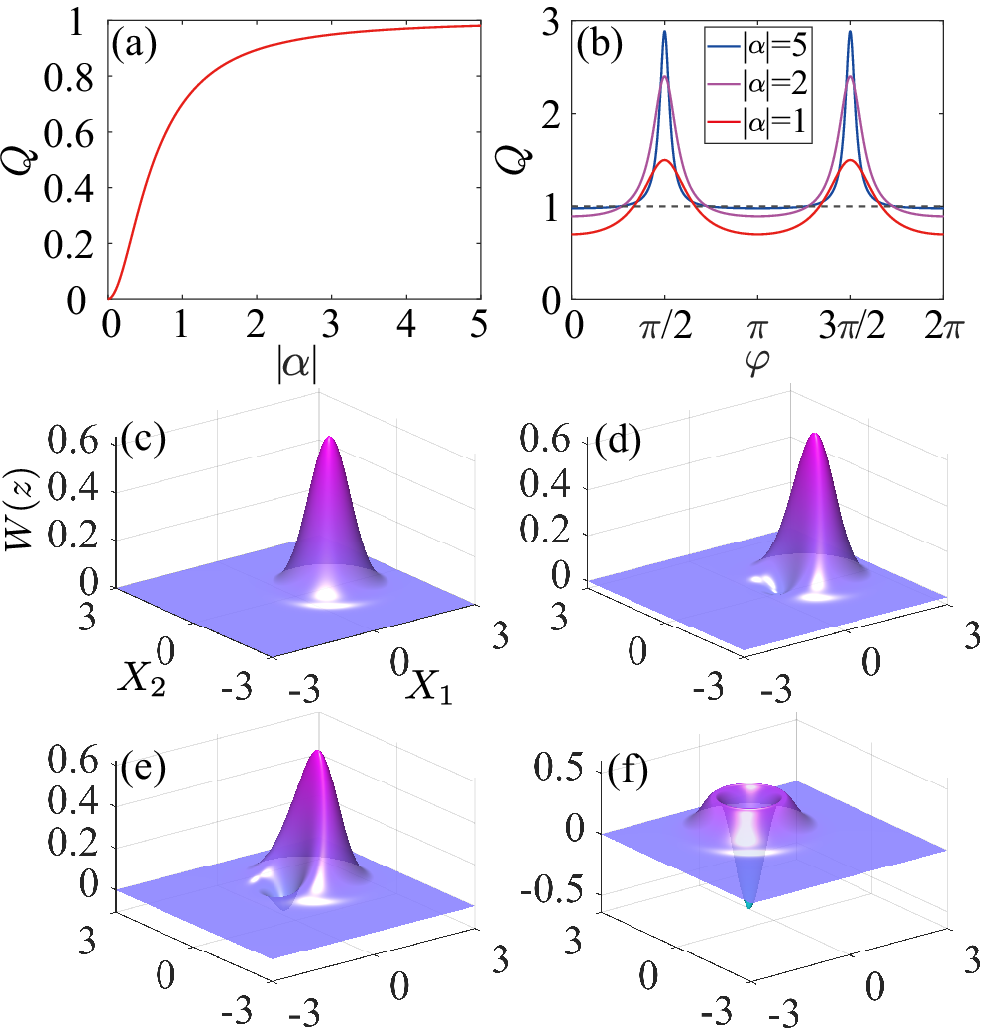}
\par\end{centering}
\caption{Nonclassicality of a quadrature-operated coherent state, i.e. the
superposition of a DFS and a coherent state, demonstrated through
$Q$-parameter and Wigner distribution. $Q$-parameter as a function
of (a) the displacement magnitude $|\alpha|$ when phase $\varphi=0$
and (b) the displacement phase $\varphi$ at various magnitudes. The
superposition retains sub-Poissonian statistics for all $|\alpha|$
at zero phase but achieves super-Poissonian at $\pi/2$ and $3\pi/2$
phases. Accordingly, a varying squeezed state is formed, depending
on the phase $\varphi$. Wigner distributions $W(z)$ over the phase
space $z=X_{1}+iX_{2}$ for (c) a coherent state with the Gaussian
distribution and quadrature-operated coherent states with displacements
(d) $\alpha=1$, (e) $\alpha=(1+i)/\sqrt{2}$, and (f) $\alpha=i$.
The nonclassicality is characterized by negative values obtained in
$W(z)$. Turning the displacement from the real to the imaginary axis,
the distribution of negative $W$ has increased until $\left|\psi\right\rangle $
reaches $\left|1^{(i)}\right\rangle $ that resembles most to a Fock
state.~\protect\label{fig:sub-Poi}}
\end{figure}

The quadrature difference operator $(q_{t}-q_{r})$ in Eq.~(\ref{eq:psi2})
has two effects: (i) each quadrature operator creates a nonclassical
superposition within its respective Hilbert space, and (ii) the difference
operation as a whole generates entanglement across the two Hilbert
spaces. By (i), we mean the quadrature operation in either $\mathcal{H}_{t}$
or $\mathcal{H}_{r}$ effectively generates the state
\begin{equation}
q\left|\alpha\right\rangle =\left(a+a^{\dagger}\right)\left|\alpha\right\rangle =2\alpha_{R}|\alpha\rangle+|1^{(\alpha)}\rangle\label{eq:quad_meas_state}
\end{equation}
from any coherent state $\left|\alpha\right\rangle $~\citep{Sesha15}
such that the resulting superposition comprises a displaced Fock state
(DFS) $\left|1^{(\alpha)}\right\rangle =D(\alpha)\left|1\right\rangle $
and the original coherent state with coefficient $\alpha_{R}=\Re\{\alpha\}$.
Since $\left\langle \alpha|1^{(\alpha)}\right\rangle =0$, the two
terms on the RHS, though not eigenstates of $q$, are orthogonal.
The nonclassicality of such a superposition is exhibited, qualitatively,
by its heralded addition of idler photons~\citep{Biagi} and, quantitatively,
by its unique photon-number variance measured by Mandel's parameter
$Q=\left\langle \Delta(a^{\dagger}a)^{2}\right\rangle /\left\langle a^{\dagger}a\right\rangle \geq0$~\citep{Agarwal-1,Mandel}.
As shown in Fig.~\ref{fig:sub-Poi}(a), $Q$ vanishes for the Fock
state $\left|1\right\rangle $ when $|\alpha|=0$ and converges to
unity for a coherent state with Poissonian distribution over the Fock
basis. For the nonclassical states, however, they may exhibit either
sub-Poissonian ($Q<1$) or super-Poissonian ($Q>1$) distributions,
depending on the displacement phase $\varphi$ as shown in Fig.~\ref{fig:sub-Poi}(b).
In such situations, the nonclassicality can be verified by the negativity
of the Wigner function in the phase space corresponding to the squeezed
or anti-squeezed photon-number variance.

For instance, the nonclassicality of a SPACS can be characterized
by a dipping towards negative axis in Wigner distributions~\citep{Zavatta}.
For the nonclassical state of (\ref{eq:quad_meas_state}), its Wigner
function being
\begin{equation}
W(z)=\frac{2}{\pi N}\left(4|z-\alpha_{I}|^{2}-1\right)e^{-2|z-\alpha|^{2}}
\end{equation}
shows similar characterizations, as shown in Fig.~\ref{fig:sub-Poi}(d)--(f).
In the superposition of $\left|\alpha\right\rangle $ and $\left|1^{(\alpha)}\right\rangle $,
the coherent-state term exemplifies when $\alpha$ has a large real
part; whereas, only the DFS term remains when $\alpha$ is purely
imaginary. Therefore, as shown in the plots, varying from $\alpha=1$
to $\alpha=i$, the Wigner function is increasingly removed from the
typical Gaussian for a coherent state towards a volcano-shaped surface
with a center dip. This change coincides with the intuitive view of
a classical-to-quantum crossover.

For effect (ii) of the difference operation over the nonclassical
states generated from quadrature operations, we mean such operation
would create entanglement between the photon statistics across the
$t$- and the $r$-branch beams. To facilitate classical nonseparability
on them, which is defined over a unified beam of orthogonal polarizations,
the $r$-branch of the $\left|\psi_{2}\right\rangle $ is $\pi/2$-rotated
by another half-wave plate (HWP2) and recombine with the $t$-branch
using a polarizing beam splitter (PBS2). The resulting beam thus become
$x$- and $y$-polarized again, whose quantum state reads
\begin{multline}
\left|\psi_{3}\right\rangle =\frac{1}{\sqrt{2N}}\Bigl[\left|1^{(\mu_{x})}\right\rangle |\mu_{y}\rangle-|\mu_{x}\rangle\left|1^{(\mu_{y})}\right\rangle \\
+2\sqrt{2}\xi_{R}|\mu_{x}\rangle|\mu_{y}\rangle\Bigr].\label{eq:psi3}
\end{multline}

Nevertheless, the displacements $\mu_{x}=(\eta+\xi)/\sqrt{2}$ and
$\mu_{y}=(\eta-\xi)/\sqrt{2}$ differ largely from the original polarization
magnitudes. Nonclassical superpositions with polarization-dependent
coefficient $\xi_{R}=\Re\{\xi\}$ and the polarization remixing between
$\xi$ and $\eta$ effects a state containing both quantifiable entanglement
and classical nonseparability. Since the coherent states are orthogonal
to both displaced Fock states in Eq.~\ref{eq:psi3}, $\left|\psi_{3}\right\rangle $
is akin to a Bell state in a two-qubit Hilbert space when $\xi_{R}$
vanishes.

State $|\psi_{3}\rangle$ is an entangled state containing both the
quantum entanglement and the classical nonseparability. The entanglement
is measured by negativity~\citep{Vidal} on a pure-state density
matrix $\rho=|\psi\rangle\langle\psi|$ through a partial transpose
$T$ on one of the subspaces. We extend negativity to the continuous
spaces to have $\mathcal{N}(\rho)=(||\rho^{T_{x}}||_{1}-1)/2$, where
the partial transpose is applied on $\mathcal{H}_{x}$ and the trace
norm $||\cdot||_{1}$ is effectively the finite sum of negative eigenvalues
of $\rho^{T_{x}}$ (Cf. Appendix B), to find 
\begin{equation}
\mathcal{N}(\rho)=\frac{1}{2+8\xi_{R}^{2}}.\label{eq:negativity}
\end{equation}
When the displacement is purely imaginary along the vertical quadrature
axis with $\xi_{R}=0$, $\mathcal{N}(\rho)$ obtains its maximal value
$1/2$, verifying our expectation that the case corresponds to a purely
quantum Bell-like $\left|\psi_{3}\right\rangle $.

\section{Mixture of entanglement and classical nonseparability}

On the other hand, the classical nonseparability, measured by Schmidt
number, has been defined through the electric field vector $\mathbf{E}$
in Euclidean space~\citep{Qian} rather than the density matrix since
the classical picture lacks the Hilbert space description. To reconcile
this conflict with the quantum interpretation, we promote the field
vector to the field operator $\hat{E}=\hat{E}_{x}+\hat{E}_{y}$ over
the two polarizations and compute Schmidt number from $\hat{E}$ as
an observable on the last step of the gedanken experiment of Fig~\ref{fig:scheme}
(indicated as a detector). Writing $\hat{E}_{x}=\mathbf{e}_{x}\mathscr{E}a_{x}\exp\{-i(\omega t-kz)\}+\mathrm{h.c.}$
and similarly for $\hat{E}_{y}$ with field amplitude unit $\mathscr{E}=\sqrt{\hbar\omega/2\varepsilon_{0}V}$
and polarization unit vector $\mathbf{e}_{x}$, the measurement expectation
of the field is
\begin{multline}
\mathbf{E}=\langle\psi_{3}|\hat{E}|\psi_{3}\rangle=\sum_{m\in\{x,y\}}\mathscr{E}\mathbf{e}_{m}\times\\
\left[|\mu_{m}|\cos(\omega t-kz-\varphi_{m})\pm r\cos(\omega t-kz)\right]\label{eq:E_field}
\end{multline}

Here, $\varphi_{m}$ denotes the phase of $\mu_{m}$, $r=\sqrt{2}\xi_{R}/(1+4\xi_{R}^{2})=\sqrt{2}\xi_{R}/N$
measures the horizontal displacements in the original coherent state$\left|\xi\right\rangle _{x}$,
and the sign of the second term is $+$ ($-$) for $x$- ($y$-) polarization
(Cf. Appendix C). The overlap of the polarization amplitudes into
one another direction is obvious in Eq.~(\ref{eq:E_field}), in addition
to the extra $r$ term which appears in $y$-polarization despite
its origin in the $x$-polarization.

Such overlap constitutes a finite classical separability as reflected
in the Schmidt number $K(\mathcal{W})=\left[1-\textrm{sin}^{2}(\Delta\phi)\textrm{sin}^{2}(2\theta)/2\right]^{-1}$
defined through a polarization matrix $\mathcal{W}$ in the lab frame~\citep{You}.
The Schmidt number, falling in the range of $1\leq K\leq2$, represents
the degree of classical nonseparability~\citep{Gabriel,Qian}. The
lower bound corresponds to a separable state, while the upper bound
indicates a Bell-like state. The mixture of polarization appears in
both the new polarization angle 
\begin{equation}
\theta=\arctan\sqrt{\frac{|\mu_{y}|^{2}-2r|\mu_{y}|\cos\varphi_{y}+r^{2}}{|\mu_{x}|^{2}+2r|\mu_{x}|\cos\varphi_{x}+r^{2}}}\label{eq:theta}
\end{equation}
and the difference $\Delta\phi=\phi_{y}-\phi_{x}$ between the new
phases 
\begin{equation}
\phi_{m}=\arctan\left(\frac{|\mu_{m}|\sin\varphi_{m}}{|\mu_{m}|\cos\varphi_{m}\pm r}\right),\label{eq:phi_m}
\end{equation}
the sign being $+\,(-)$ for $x$-($y$-) polarization. At $r=0$,
$K(\mathcal{W})$ obtains its maximal value when $\mu_{y}=\pm i\mu_{x}$,
i.e. when $\xi_{R}=\eta_{I}$ and $\xi_{I}=\pm\eta_{R}$; the case
where the $y$-polarization is $\pi/2$ ahead (behind) the $x$-polarization
or the beam at $\left|\psi_{0}\right\rangle $ is CCW (CW) circularly
polarized. If $r\neq0$, one of the scenario for maximal $K(\mathcal{W})$
associates with the conditions $\mu_{y}=-\mu_{x}^{\ast}$, i.e. when
$\xi_{R}/\xi_{I}=-\eta_{I}/\eta_{R}$. This case includes the previous
scenario as a subset and other scenarios of disproportioned amplitudes
among $\xi$ and $\eta$, the latter of which correspond to beams
with elliptical polarization.

\begin{table}
\begin{centering}
\begin{tabular}{|c||>{\centering}p{2cm}|>{\centering}p{2cm}|}
\hline 
 & $\mathcal{N}(\rho)$ max & $\mathcal{N}(\rho)$ min\tabularnewline
\hline 
\hline 
$K(\mathcal{W})$ max & $\xi_{R}=\eta_{I}=0$

$\xi_{I}=\pm\eta_{R}$ & $\xi_{R}\to\infty$

$\eta=\pm i\xi$\tabularnewline
\hline 
$K(\mathcal{W})$ min & $\xi_{R}=\eta_{R}=0$

$\xi_{I}=\pm k\eta_{I}$ & $\xi_{R}\to\infty$

$\eta=\pm k\xi$\tabularnewline
\hline 
\end{tabular}
\par\end{centering}
\caption{Conditions among the beam $x$- and $y$-polarizations $\xi=\xi_{R}+i\xi_{I}$
and $\eta=\eta_{R}+i\eta_{I}$ for obtaining the extrema of the negativity
$\mathcal{N}(\rho)$ and the Schmidt number $K(\mathcal{W})$, where
$r$ is set to zero. $k$ is a real proportional constant.}

\end{table}

Since the negativity depends on the polarization amplitude $\xi_{R}$
while the Schmidt number depends on the scaled $r$, their extrema
do not coincide but are obtainable independently despite the functional
dependence of $r$ on $\xi_{R}$. The entanglement and the classical
nonseparability can be separately maximized or minimized depending
on the input beam polarization, as shown in Table~I, where the only
common constraint is $r=0$. In particular, the simultaneity of maximal
negativity~(\ref{eq:negativity}) and Schmidt number occurs for a
circularly polarized beam with zero $\xi_{R}$ in its $x$-polarization.
Conversely, a linearly polarized beam with proportional $x$- and
$y$- polarizations such that $\xi_{R}=\eta_{R}$ is sufficiently
large generates the simultaneous minimum. For the latter, the large
$\xi_{R}$ and $\eta_{R}$ limit corresponds to vanishing $\mu_{y}$
and $r$, leading the $\mathbf{E}$ field of Eq.~(\ref{eq:E_field})
back to a unmixed classical state containing only $x$-polarization.

\begin{figure}
\begin{centering}
\includegraphics[clip,width=8.6cm]{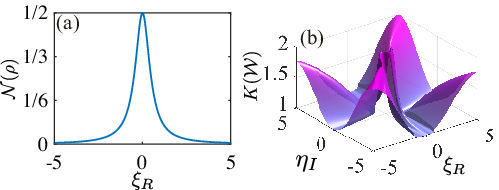}
\par\end{centering}
\caption{(a) Negativity for measuring the quantum entanglement and (b) Schmidt
number for measuring the classical nonseparability. Both measures
the nonclassical state $\left|\psi_{3}\right\rangle $, and for (b)
$\eta_{R}=1$ and $\xi_{I}=-1$.~\protect\label{fig:neg_Schmidt}}
\end{figure}

Figures \ref{fig:neg_Schmidt}(a)--(b) show, respectively, negativity
and Schmidt number as functions of $\xi_{R}$ and $\eta_{I}$. Negativity
is symmetric about the $\xi_{R}=0$ axis. Since the quadrature $q$
only measures $x$-polarized component along light path 1 in Fig.~(\ref{fig:scheme}),
it verifies that the quantum entanglement is solely associated with
vacuum or displaced Fock states. Because of its asymmetric placement
between paths 1 and 2, the quadrature operation further breaks the
mirror symmetry about the $\xi_{R}=0$ plane in the Schmidt number
that measures the classical nonseparability. Rather, the mirror symmetry
arises on the diagonal $\xi_{R}=\pm\eta_{I}$ planes when the operation
on the state $\left|\psi_{1}\right\rangle $ has mitigated effects
for large $\xi_{R}$ and the two light paths have symmetric polarization
magnitudes.

\section{Experimental proposal for state detection}

\begin{figure}
\begin{centering}
\includegraphics[clip,width=8.6cm]{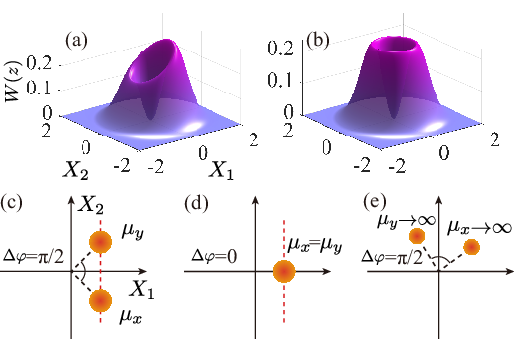}
\par\end{centering}
\caption{Wigner distributions about the reduced density matrix $\rho_{x}$
(associated with $x$-polarization of the output light beam at state$\left|\psi_{3}\right\rangle $).
Surface plots for small displacements $\mu_{x}=(\xi+\eta)/\sqrt{2}=0.1$
are shown in (a) where $\sqrt{2}\xi_{R}=0.1$ and in (b) where $\sqrt{2}\xi_{R}=0$.
The slight difference in $\xi_{R}$ incurs a symmetry breaking about
the center. Contour plots for large displacements are shown in (c)$\text{\textendash}$(e):
(c) shows the case where entanglement and classical nonseparability
are both maximal; (d) shows the case with maximal entanglement and
vanishing classical nonseparability; and (e) shows the case with maximal
classical nonseparability and vanishing entanglement.~\protect\label{fig:Wigner_entangled}}
\end{figure}

As illustrated in Fig.~\ref{fig:sub-Poi} above, the nonclassical
superposition state $\left|\psi_{2}\right\rangle $ is uniquely described
by a Wigner distribution $W(z)$~\citep{Zavatta,Lvovsky-1,Breitenbach}.
The entangled $\left|\psi_{3}\right\rangle $ containing a mixture
of entanglement and classical nonseparability over the Hilbert spaces
$\mathcal{H}_{x}$ and $\mathcal{H}_{y}$ can then be detected using
Wigner distributions along both polarization directions. The character
of the mixture is readily identified by the unique locations and shapes
derived from tomographic measurement on either polarization, where
the other polarization is traced out over the orthogonal basis $\{\left|\mu_{m}\right\rangle ,\left|1^{(\mu_{m})}\right\rangle \}$.
For instance, if we measure the density matrix $\rho_{x}=\mathrm{tr}_{y}\{\left|\psi_{3}\right\rangle \left\langle \psi_{3}\right|\}$
on the $x$-polarization, its Wigner distribution reads
\begin{equation}
W(z)=\frac{4|z-\mu_{x}+\sqrt{2}\xi_{R}|^{2}}{(1+4\xi_{R}^{2})\pi}e^{-2|z-\mu_{x}|^{2}},
\end{equation}
for which the circular symmetry about $z=\mu_{x}$ is broken by the
addition of $\sqrt{2}\xi_{R}$ in the quadratic factor but not in
the Gaussian, echoing the effect we already observed in Fig.~\ref{fig:neg_Schmidt}(b).
Geometrically, the symmetry breaking corresponds to a transition from
a volcano shape to one with a slanted crest, as shown in Fig.~\ref{fig:Wigner_entangled}(a)--(b).

Removed from the origin, the relative locations of the displacement
of $\mu_{x}$ and $\mu_{y}$ on the quadrature plane determine the
degrees of both entanglement and classical nonseparability convertible
from $\left|\mu_{x}\right\rangle \left|\mu_{y}\right\rangle $. For
instance, as shown in Fig.~\ref{fig:Wigner_entangled}(c)--(e),
the case of simultaneous maximal entanglement and classical nonseparability
occurs for $\mu_{x}$ and $\mu_{y}$ symmetric about the horizontal
quadrature axis, i.e. they having the same real part but non-zero
opposite imaginary parts. When the imaginary part vanishes and hence
the displacements coincide, only the entanglement survives. All states
with coinciding displacements are purely entangled states (without
classical nonseparability) though this is not true vice versa as inferred
from Eqs.~(\ref{eq:theta})--(\ref{eq:phi_m}). When the displacements
become large and differ by a phase of $\pi/2$, the classical nonseparability
survives and the entanglement vanishes.

\begin{figure}
\begin{centering}
\includegraphics[width=8.6cm]{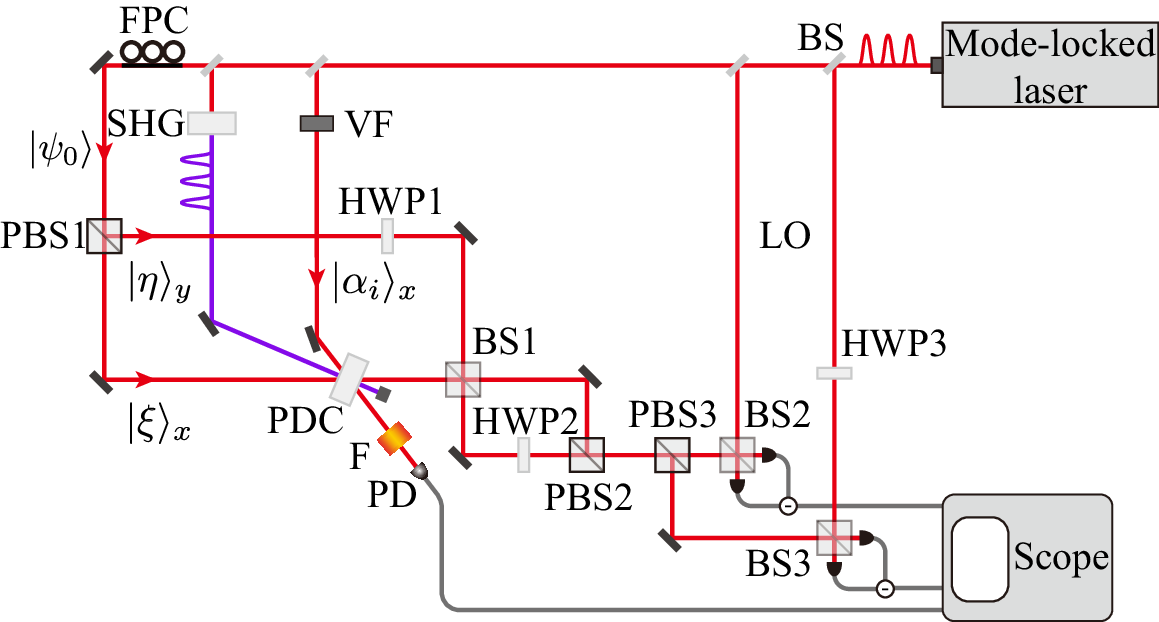}
\par\end{centering}
\caption{Proposed experimental setup for detecting entangled and classically
nonseparable states converted from nonclassical states. The components
used include: beam splitters (BS), polarized beam splitters (PBS),
half-wave plates (HWP), variable filters (VF), parametric down converter
(PDC), second harmonic generator (SHG), and photon detectors (PD).~\protect\label{fig:setup}}
\end{figure}

To demonstrate the varying degrees of entanglement and classical nonseparability,
we consider the experimental setup shown in Fig.~\ref{fig:setup}.
The single-mode light source is generated from a picosecond pulsed
mode-locked laser, which is split into two paths by a BS: one is used
as the local oscillator (LO) input for the homodyne detection later
on and the other is further split into three paths. The pulses along
the first path is attenuated by a variable filter (VF) to act as the
weak idler $|\alpha_{i}\rangle_{x}\approx|0\rangle_{x}+\alpha_{i}|1\rangle_{x}$.
The second path is fed to a nonlinear crystal for second harmonic
generation (SHG), generating the pump pulses for the PDC. The third
path goes through a fiber polarization controller (FPC) to generate
an arbitrarily polarized initial state $|\psi_{0}\rangle$ . Through
the subsequent polarizing beam splitter, $\left|\psi_{0}\right\rangle $
is split into the two orthogonally polarized states $|\xi\rangle_{x}$
and $|\eta\rangle_{y}$. The half-wave plate HWP1 rotates $|\eta\rangle_{y}$
into $\left|\eta\right\rangle _{x}$ such that the two polarizing
paths interfere with each other at BS1.

Before the interference at BS1, $\left|\xi\right\rangle _{x}$ acting
as the signal is mixed with the idler $\left|\alpha_{i}\right\rangle _{x}$
and the pump at a nonlinear crystal for the PDC process. For simplicity,
we omit the polarization subscript and denote the input signal and
idler by the product state $\left|\xi\right\rangle _{1}(|0\rangle_{i}+\alpha_{i}|1\rangle_{i})$.
The quadrature operation in Fig.~\ref{fig:scheme} is conditioned
on single-photon measurements of the idler output at PD after the
three-wave mixing. The mixing approximately has the effect of $1+ga_{1}^{\dagger}a_{i}^{\dagger}+g^{*}a_{1}a_{i}$
on the input state $\left|\xi\right\rangle _{1}(|0\rangle_{i}+\alpha_{i}|1\rangle_{i})$
when the parametric gain $g$ is sufficiently low ($|g|\ll1$)~\citep{Zavatta},
giving the output
\begin{equation}
(1+g^{*}\xi\alpha_{i})\left|\xi\right\rangle _{1}|0\rangle_{i}+(\alpha_{i}+ga_{1}^{\dagger})\left|\xi\right\rangle _{1}|1\rangle_{i}+\sqrt{2}g\alpha_{i}a_{1}^{\dagger}\left|\xi\right\rangle _{1}|2\rangle_{i}.
\end{equation}
When one photon is recorded by the PD, the signal state is projected
onto $g(\alpha_{i}/g+a_{1}^{\dagger})|\xi\rangle_{1}$.\textbf{ }Letting
$\alpha_{i}=g\xi$, the emitted signal state would be equivalent to
the desired $g(a_{1}+a_{1}^{\dagger})|\xi\rangle_{1}$ and the normalized
system state would become $|\psi_{1}\rangle$.

The 50-50 beam splitting at BS1 then generates $\left|\psi_{2}\right\rangle $
as in Fig.~\ref{fig:scheme} along two legs, which are spatially
recombined by PBS2 after the reflection leg is rotated by HWP2. Hence,
the final state $\left|\psi_{3}\right\rangle $ containing complex
amplitudes along both the $x$- and $y$-polarizations is prepared.
The beam splittings at BS2 and BS3 separate the quadratures for each
polarization direction, which are individually detected for their
time correlations. The relative phase between the LO and the signal
can be adjusted by a piezoelectric transducer (not shown) in the LO.
The measurement results of PD after spectral and spatial filters (F),
which herald the preparation of the nonclassical state, are used to
select the results from balanced homodyne detection.

\section{Conclusions and discussions}

We demonstrate the convertibility of a classical polarized beam to
an macroscopic entangled state, where the relationship between the
degrees of quantum entanglement and classical nonseparability obtained
from the polarizations of the beam is established. The convertibility
is realized by the nonclassicality obtainable from a coherent single-mode
beam through linear optics and, by inserting a quadrature operation
asymmetrically along one polarization path, arbitrary mixtures of
entanglement and classical nonseparability are eventually converted
from an appropriate pair of displacements from the beam polarizations.

In other words, we have shown full range of quantum entanglement and
classical nonseparability can be simultaneously generated from macroscopic
polarizations. The computation method provides a means to characterize
both the quantum and classical aspects of a single light state.\textcolor{blue}{{}
}The convertibility from polarizations here is useful for developing
state preparation and quantum information processing techniques that
take advantage of both the unique property of entanglement and the
ease of operation of classical beams. Our proposed experimental setup
demonstrates its viability.\textcolor{red}{{} }
\begin{acknowledgments}
H. I. thanks the support of FDCT Macau under grants 006/2022/ALC and
0179/2023/RIA3.
\end{acknowledgments}

\appendix

\section{Quantum description of polarization conversion}

The initial state of the system is described by the product state
\begin{equation}
|\psi_{0}\rangle=\left|\xi\right\rangle _{x}\left|\eta\right\rangle _{y},
\end{equation}
where $\xi$ and $\eta$ parametrize the coherent states in two independent
polarization directions $x$ and $y$, respectively. 

In path 1, the state $|\xi\rangle_{1}$ undergoes the conditional
quantum operation $q=a+a^{\dagger}$, generating the nonclassical
superposition state $(a_{1}+a_{1}^{\dagger})|\xi\rangle_{1}$. Therefore,
the normalized state $|\psi_{1}\rangle$ reads 
\begin{equation}
|\psi_{1}\rangle=\frac{1}{\sqrt{N}}\left(a_{1}+a_{1}^{\dagger}\right)\left|\xi\right\rangle _{1}\left|\eta\right\rangle _{2}.
\end{equation}

The operation enacted by the 50-50 beam splitter is represented by
the superoperator $\mathcal{B}=\exp\{(a_{1}^{\dagger}a_{2}-a_{1}a_{2}^{\dagger})\pi/4\}$.
The annihilation operators $a_{1}$ and $a_{2}$ are transformed through
the unitary transformations
\begin{align}
\mathcal{B}a_{1}\mathcal{B}^{\dagger} & =a_{1}\textrm{cos}\frac{\pi}{4}-a_{2}\textrm{sin}\frac{\pi}{4}=\frac{1}{\sqrt{2}}\left(a_{t}-a_{r}\right),\nonumber \\
\mathcal{B}a_{2}\mathcal{B}^{\dagger} & =a_{2}\textrm{cos}\frac{\pi}{4}+a_{1}\textrm{sin}\frac{\pi}{4}=\frac{1}{\sqrt{2}}\left(a_{r}+a_{t}\right),\label{eq:beam=000020splitter}
\end{align}
where we have used the Baker-Hausdorff formula and renamed the operators
$a_{1}$ and $a_{2}$ to $a_{t}$ and $a_{r}$, respectively, to indicate
the transition from input fields to output fields.

Letting $\mathcal{B}$ operate on the state $|\psi_{1}\rangle$ and
utilizing the property $\mathcal{B}^{\dagger}\mathcal{B}=1$, we obtain
the state $|\psi_{2}\rangle$ as
\begin{align}
\mathcal{B}|\psi_{1}\rangle & =\frac{1}{\sqrt{N}}\mathcal{\mathcal{B}}\left(a_{1}+a_{1}^{\dagger}\right)D_{1}(\xi)D_{2}(\eta)|0\rangle_{1}|0\rangle_{2}\nonumber \\
 & =\frac{1}{\sqrt{N}}\mathcal{\mathcal{B}}\left(a_{1}+a_{1}^{\dagger}\right)\mathcal{B}^{\dagger}\mathcal{B}D_{1}(\xi)\mathcal{B}^{\dagger}\mathcal{B}D_{2}(\eta)\mathcal{B}^{\dagger}\mathcal{B}|0\rangle_{1}|0\rangle_{2}.
\end{align}
The effect of $\mathcal{B}$ on the quadrature operator $q_{1}=a_{1}+a_{1}^{\dagger}$
can be seen by using the Eq.~\ref{eq:beam=000020splitter}. The displacement
operators $D_{1}$ and $D_{2}$ are transformed through the formulae
$\mathcal{B}D_{1}(\xi)\mathcal{B}^{\dagger}=D_{t}(\xi/\sqrt{2})D_{r}(-\xi/\sqrt{2})$
and $\mathcal{B}D_{2}(\eta)\mathcal{B}^{\dagger}=D_{t}(\eta/\sqrt{2})D_{r}(\eta/\sqrt{2})$.
Moreover, the beam splitter has no effect on the vacuum, i.e. we can
write $\mathcal{B}|0\rangle_{1}|0\rangle_{2}=|0\rangle_{t}|0\rangle_{r}$.
Hence, we have
\begin{align}
|\psi_{2}\rangle= & \frac{a_{t}-a_{r}+a_{t}^{\dagger}-a_{r}^{\dagger}}{\sqrt{2N}}D_{t}\left(\frac{\eta+\xi}{\sqrt{2}}\right)D_{r}\left(\frac{\eta-\xi}{\sqrt{2}}\right)|0\rangle_{t}|0\rangle_{r}\nonumber \\
= & \frac{1}{\sqrt{2N}}(q_{t}-q_{r})\left|\frac{\eta+\xi}{\sqrt{2}}\right\rangle _{t}\left|\frac{\eta-\xi}{\sqrt{2}}\right\rangle _{r}
\end{align}
as an entangled state regarding the transmission and reflection modes.
After the polarization rotation of path 2 and recombination by PBS2,
the final state $|\psi_{3}\rangle$ is obtained. With the commutation
relations 
\begin{align}
\left[a,D(\alpha)\right] & =\alpha D(\alpha),\nonumber \\
\left[a^{\dagger},D(\alpha)\right] & =\alpha^{*}D(\alpha),
\end{align}
we can derive 
\begin{align}
q|\alpha\rangle & =\left[D(\alpha)a+\alpha D(\alpha)+D(\alpha)a^{\dagger}+\alpha^{*}D(\alpha)\right]|0\rangle\nonumber \\
 & =|1^{(\alpha)}\rangle+2\alpha_{R}|\alpha\rangle.
\end{align}
Therefore, the state $\left|\psi_{3}\right\rangle $ is represented
by orthogonal states in the form shown in Eq.~\ref{eq:psi3}.

The Wigner function, serving as a quasi-probability distribution of
a quantum state, is derived by calculating an integral. For example,
the Wigner function of the nonclassical superposition state $(a+a^{\dagger})|\alpha\rangle=q|\alpha\rangle$
is calculated as follows 
\begin{align}
W(z) & =\frac{2e^{2|z|^{2}}}{\pi^{2}N}\int d^{2}\beta\left\langle -\beta|q|\alpha\right\rangle \left\langle \alpha|q^{\dagger}|\beta\right\rangle e^{2(\beta^{*}z-\beta z^{*})}\nonumber \\
 & =\frac{2e^{2|z|^{2}-|\alpha|^{2}}}{\pi N}(|2z-\alpha+\alpha^{*}|^{2}-1)e^{-|2z-\alpha|^{2}}\nonumber \\
 & =\frac{2}{\pi N}(4|z-\alpha_{I}|^{2}-1)e^{-2|z-\alpha|^{2}}.
\end{align}

\section{Negativity from a nonclassical state}

According to the Peres-Horodecki separability criterion~\citep{Peres,Horodecki},
the separability for bipartite qubit systems is characterized by the
non-negativeness of the eigenvalues of the joint system density matrix.
Vidal and Werner~\citep{Vidal} quantify this characteristic in eigenvalues
through negativity as the entanglement measure. Taking the standard
Fock-state expansion of a coherent state in a countably infinite dimensional
space~\citep{Glauber63}, the negativity is naturally extended to
coherent states $\left|\mu\right\rangle $ and displaced Fock states
$\left|1^{(\mu)}\right\rangle $. In other words, due to the orthogonality
$\left\langle \mu|1^{(\mu)}\right\rangle =0$, the pair $\left|\mu\right\rangle $
and $\left|1^{(\mu)}\right\rangle $ constitute the eigenbasis of
a Bloch sphere displaced from the origin of the quadrature plane by
the amount of $\mu$, just as they do for the Bloch sphere without
displacement, i.e. the conventional $\left|0\right\rangle $ and $\left|1\right\rangle $
Fock states when $\mu\to0$. Since the displacement amounts $\mu_{x}$
and $\mu_{y}$ remain committed in their respective Hilbert space
without intermixing, the computation of entanglement through negativity
is not affected by the change of basis because of the displacements.

Therefore, for the final state $\left|\psi_{3}\right\rangle $, we
have
\begin{align}
\rho= & \frac{1}{2N}\Bigl\{\left|1^{(\mu_{x})},\mu_{y}\right\rangle \left\langle 1^{(\mu_{x})},\mu_{y}\right|+\left|\mu_{x},1^{(\mu_{y})}\right\rangle \left\langle \mu_{x},1^{(\mu_{y})}\right|\nonumber \\
+ & \Bigl[2\sqrt{2}\xi_{R}\left(\left|1^{(\mu_{x})},\mu_{y}\right\rangle -\left|\mu_{x},1^{(\mu_{y})}\right\rangle \right)\left\langle \mu_{x},\mu_{y}\right|\nonumber \\
- & \left|1^{(\mu_{x})},\mu_{y}\right\rangle \left\langle \mu_{x},1^{(\mu_{y})}\right|+\mathrm{h.c.}\Bigr]+8\xi_{R}^{2}\left|\mu_{x},\mu_{y}\right\rangle \left\langle \mu_{x},\mu_{y}\right|\Bigr\}
\end{align}
where we have used the abbreviation such as $\left|1^{(\mu_{x})},\mu_{y}\right\rangle \left\langle 1^{(\mu_{x})},\mu_{y}\right|=\left|1^{(\mu_{x})}\right\rangle \left\langle 1^{(\mu_{x})}\right|\otimes\left|\mu_{y}\right\rangle \left\langle \mu_{y}\right|$
without confusion and let $\xi_{R}=\Re\{\mu_{x}-\mu_{y}\}/\sqrt{2}$
denote the real part of $\xi$. The partial transpose $T_{x}$ has
the effect of $\left|m\right\rangle \left\langle n\right|\to\left|n\right\rangle \left\langle m\right|$
on the first subspace $\mathcal{H}_{x}$ under the Fock-state basis.
Applying the effect on the expansions of the outer product of two
coherent states, we find
\begin{align}
\left|\alpha\right\rangle \left\langle \beta\right| & =\exp\left\{ -\frac{|\alpha|^{2}+|\beta|^{2}}{2}\right\} \sum_{m,n}\frac{\alpha^{m}\beta^{\ast n}}{\sqrt{m!n!}}\left|m\right\rangle \left\langle n\right|\nonumber \\
 & \to\exp\left\{ -\frac{|\alpha|^{2}+|\beta|^{2}}{2}\right\} \sum_{n,m}\frac{\beta^{\ast n}(\alpha^{\ast})^{\ast m}}{\sqrt{n!m!}}\left|n\right\rangle \left\langle m\right|\nonumber \\
 & =\left|\beta^{\ast}\right\rangle \left\langle \alpha^{\ast}\right|.
\end{align}
For a displaced Fock state $\left|1^{(\alpha)}\right\rangle =D(\alpha)\left|1\right\rangle $,
its Hermitian conjugate is $\left\langle 1\right|D^{\dagger}(a)=\left\langle 1\right|e^{\alpha^{\ast}a-\alpha a^{\dagger}}$
which can be expanded in the Fock basis $\left\{ \left\langle 0\right|,\left\langle 1\right|,\left\langle 2\right|,\dots\right\} $.
In the matrix representation, Hermitian conjugate is the combination
of transpose and complex conjugation. Thus, transpose has the effect
of Hermitian transformation with the complex conjugation stripped,
i.e. retaining the transforms $a\to a^{\dagger}$ and $a^{\dagger}\to a$
on the Fock basis but not the $\alpha\to\alpha^{\ast}$. Therefore,
$\left|1^{(\alpha)}\right\rangle ^{T}=\left\langle 1\right|e^{\alpha a-\alpha^{\ast}a^{\dagger}}=\left\langle 1\right|D^{\dagger}(\alpha^{\ast})=\left\langle 1^{(\alpha^{\ast})}\right|$.
Consequently, the partially transposed density matrix is written as

\begin{align}
\rho^{T_{x}}= & \frac{1}{2N}\Bigl\{\left|1^{(\mu_{x}^{\ast})},\mu_{y}\right\rangle \left\langle 1^{(\mu_{x}^{\ast})},\mu_{y}\right|+\left|\mu_{x}^{\ast},1^{(\mu_{y})}\right\rangle \left\langle \mu_{x}^{\ast},1^{(\mu_{y})}\right|\nonumber \\
+\Bigl[2 & \sqrt{2}\xi_{R}\left(\left|\mu_{x}^{\ast},\mu_{y}\right\rangle \left\langle 1^{(\mu_{x}^{\ast})},\mu_{y}\right|-\left|\mu_{x}^{\ast},1^{(\mu_{y})}\right\rangle \left\langle \mu_{x}^{\ast},\mu_{y}\right|\right)\nonumber \\
-|\mu_{x}^{\ast} & ,\mu_{y}\rangle\left\langle 1^{(\mu_{x}^{\ast})},1^{(\mu_{y})}\right|+\mathrm{t.c.}\Bigr]+8\xi_{R}^{2}\left|\mu_{x}^{\ast},\mu_{y}\right\rangle \left\langle \mu_{x}^{\ast},\mu_{y}\right|\Bigr\}.
\end{align}

Given fixed $\mu_{x}$ and $\mu_{y}$, we can express $\rho^{T_{x}}$
in a symmetric matrix form in the orthogonal basis $\left\{ \left|1^{(\mu_{x}^{\ast})},1^{(\mu_{y})}\right\rangle ,\left|1^{(\mu_{x}^{\ast})},\mu_{y}\right\rangle ,\left|\mu_{x}^{\ast},1^{(\mu_{y})}\right\rangle ,\left|\mu_{x}^{\ast},\mu_{y}\right\rangle \right\} $:

\begin{equation}
\rho^{T_{x}}=\frac{1}{2N}\left[\begin{array}{cccc}
0 & 0 & 0 & -1\\
0 & 1 & 0 & 2\sqrt{2}\xi_{R}\\
0 & 0 & 1 & -2\sqrt{2}\xi_{R}\\
-1 & 2\sqrt{2}\xi_{R} & -2\sqrt{2}\xi_{R} & 8\xi_{R}^{2}
\end{array}\right].\label{eq:rho_matrix}
\end{equation}

\noindent Its trace norm~\citep{Reed} defined as $||\rho^{T_{x}}||_{1}=\textrm{tr}\sqrt{(\rho^{T_{x}})^{T}\rho^{T_{x}}}$
is related to the negative matrix eigenvalues and the negativity through
the equation $||\rho^{T_{x}}||_{1}=1+2|\Sigma_{i}\lambda_{i}^{-}|=1+2\mathcal{N}(\rho)$~\citep{Vidal}.
The eigenvalues of Matrix~(\ref{eq:rho_matrix}) are $\pm1/(2+8\xi_{R}^{2})$
and $1/2\pm(1/2)\sqrt{1-(1+4\xi_{R}^{2})^{-2}}$. Since the latter
two are always positive, there is a unique negative eigenvalue, letting
the negativity of $\rho$ be

\begin{equation}
\mathcal{N}(\rho)=\frac{1}{2+8\xi_{R}^{2}}.
\end{equation}

\section{Schmidt number from a nonclassical state}

The classical nonseparability measured by the Schmidt number is determined
by the two orthogonal polarization components of the electric field.
When two orthogonal polarization directions are chosen, the correlation
between the two complex amplitudes is $C_{xy}=\langle E_{x}E_{y}^{*}\rangle=|E_{x}||E_{y}|e^{i(\varphi_{x}-\varphi_{y})}$.
This function contains the information of the magnitudes and phase
difference along two orthogonal directions. After normalizing by the
light intensity $I=|E_{x}|^{2}+|E_{y}|^{2}$, the function becomes
$C_{xy}=\textrm{cos}\theta\textrm{sin}\theta e^{i(\varphi_{x}-\varphi_{y})}$.
This correlation function relates to the Schmidt number through the
relation 
\begin{equation}
K=1/\left[1-2\left(\Im\left\{ C_{xy}\right\} \right)^{2}\right].
\end{equation}
In other words, the two orthogonal polarizations are bounded by correlations
embodied by the Stokes parameters in classical optics. Equivalently,
this correlation occurs between the polarization amplitude and the
polarization direction in the one electric field, which is embodied
by the coherence-polarization matrix \citep{Gori}. Furthermore, one
can define a correlation function $C(\theta_{P},\phi_{P})$ to measure
the correlation between the polarization angle $\theta_{P}$ and the
initial phase $\phi_{P}$ of one electric field \citep{Kagalwala}.
To quantify the classical nonseparability by Schmidt number, we firstly
consider the annihilation operator $a_{x}$ in the field operator
$\hat{E}_{x}$. It is an eigen-operator of the coherent state $\left|\mu_{x}\right\rangle $,
i.e. $a_{x}\left|\mu_{x}\right\rangle =\mu_{x}\left|\mu_{x}\right\rangle $,
and has the effect of $a_{x}\left|1^{(\mu_{x})}\right\rangle =D(\mu_{x})(a_{x}+\mu_{x})\left|1_{x}\right\rangle $
on a DFS. The associated expectations are 
\begin{align}
\left\langle \mu_{x}\right|a_{x}\left|\mu_{x}\right\rangle  & =\left\langle 1^{(\mu_{x})}\right|a_{x}\left|1^{(\mu_{x})}\right\rangle =\mu_{x},\nonumber \\
\left\langle 1^{(\mu_{x})}|a_{x}|\mu_{x}\right\rangle  & =0,\label{eq:elements-1}\\
\left\langle \mu_{x}|a_{x}|1^{(\mu_{x})}\right\rangle  & =1.\nonumber 
\end{align}
Then, given the superposition~(4) in the text, one finds $\left\langle \psi_{3}|a_{x}|\psi_{3}\right\rangle =(\mu_{x}+\sqrt{2}\xi_{R}+4\mu_{x}\xi_{R}^{2})/N$.
Since the three states on the RHS of Eq.~(4) are orthonormal, the
normalization constant $N$ can be determined and one has
\begin{equation}
\left\langle \psi_{3}|a_{x}|\psi_{3}\right\rangle =\mu_{x}+r
\end{equation}
where $r=\sqrt{2}\xi_{R}/(1+4\xi_{R}^{2})=\sqrt{2}\Re\{\xi\}/N$ signifies
the scaled $\xi$-displacement. For the conjugate creation operator
$a_{x}^{\dagger}$, the first line of Eq.~(\ref{eq:elements-1})
still holds while the RHS for the second and the third lines switch.
Therefore, the field operator has the state average
\begin{multline}
\left\langle \psi_{3}\right|\hat{E}_{x}\left|\psi_{3}\right\rangle =\mathbf{e}_{x}\mathscr{E}\left[(\mu_{x}+r)e^{-i(\omega t-kz)}+\mathrm{h.c.}\right]\\
=2\mathbf{e}_{x}\mathscr{E}\left[|\mu_{x}|\cos(\omega t-kz-\varphi_{x})+r\cos(\omega t-kz)\right],\label{eq:Ex_avg-1}
\end{multline}
where $\varphi_{x}$ is the phase of $\mu_{x}$. Subsequently for
$\hat{E}_{y}$, since $\left|\psi_{3}\right\rangle $ is anti-symmetric
about the $xy$-symmetry, Eq.~(\ref{eq:Ex_avg-1}) is modified to
\begin{multline}
\left\langle \psi_{3}\right|\hat{E}_{y}\left|\psi_{3}\right\rangle =2\mathbf{e}_{y}\mathscr{E}\Bigl[|\mu_{y}|\cos(\omega t-kz-\varphi_{y})\\
-r\cos(\omega t-kz)\Bigr]\label{eq:Ey_avg-1}
\end{multline}
and combining the two equations give Eq.~(6) of the text.

The Schmidt number $K$ is computed from the polarization matrix $\mathcal{W}=C^{T}C/I$
in the lab frame~\citep{You}, where $C$ is the coefficient matrix
\begin{equation}
\left[\begin{array}{cc}
|\mu_{x}|\textrm{cos}\varphi_{x}+r & |\mu_{x}|\textrm{sin}\varphi_{x}\\
|\mu_{y}|\textrm{cos}\varphi_{y}-r & |\mu_{y}|\textrm{sin}\varphi_{y}
\end{array}\right]
\end{equation}
extracted from the quadrature amplitudes of the carriers $\cos(wt-kz)$
and $\sin(\omega t-kz)$ of Eqs.~(\ref{eq:Ex_avg-1})--(\ref{eq:Ey_avg-1})
and $I$ is the normalizing intensity $\left\langle \hat{E}_{x}\right\rangle ^{2}+\left\langle \hat{E}_{y}\right\rangle ^{2}$
of the light beam. $\mathcal{W}$ then reads 
\begin{equation}
\left[\begin{array}{cc}
\mathrm{cos}^{2}\theta & \frac{1}{2}\mathrm{cos}\Delta\varphi\mathrm{sin}2\theta\\
\frac{1}{2}\mathrm{cos}\Delta\varphi\mathrm{sin}2\theta & \mathrm{sin}^{2}\theta
\end{array}\right]\label{eq:W_mat-1}
\end{equation}
where the angles are defined through
\begin{align}
\tan\theta & =\sqrt{\frac{|\mu_{y}|^{2}-2r|\mu_{y}|\cos\varphi_{y}+r^{2}}{|\mu_{x}|^{2}+2r|\mu_{x}|\cos\varphi_{x}+r^{2}}},\\
\Delta\phi & =\phi_{y}-\phi_{x}\\
 & =\arctan\frac{|\mu_{y}|\sin\varphi_{y}}{|\mu_{y}|\cos\varphi_{y}-r}-\arctan\frac{|\mu_{x}|\sin\varphi_{x}}{|\mu_{x}|\cos\varphi_{x}+r}.
\end{align}
By definition, $K$ is $1/\sum\lambda^{2}$ where $\lambda$ are the
eigenvalues of $\mathcal{W}$. Hence, from Eq.~(\ref{eq:W_mat-1}),
$\lambda_{\pm}=1/2\left[1\pm\sqrt{1-\sin^{2}\Delta\phi\sin^{2}2\theta}\right]$
and Eq.~(7) of the text gives the Schmidt number.


\begin{thebibliography}{99}
\bibitem{Notation}To simplify and unify the notations, we use $\left|\alpha^{(n)}\right\rangle $
to denote photon-added coherent states (a.k.a. Agarwal states) and
$\left|n^{(\alpha)}\right\rangle $ to denote displaced Fock states
throughout the paper.
\end{thebibliography}
\end{document}